\documentclass[prb, twocolumn, amsmath, showpacs]{revtex4-1}
\usepackage{graphicx}
\usepackage{amssymb}
\usepackage[latin1]{inputenc}

\begin{document}


\title{Transport through quantum spin Hall insulator/metal junctions in graphene ribbons}

\author{Elsa~Prada}
\affiliation{Instituto de Ciencia de Materiales de Madrid, (CSIC), Cantoblanco, 28049 Madrid, Spain}
\author{Georgo~Metalidis}
\affiliation{Institut f\"ur Theoretische Festk\"orperphysik
and DFG-Center for Functional Nanostructures,
Karlsruhe Institute of Technology (KIT), D-76128 Karlsruhe, Germany}

\date{\today}

\begin{abstract}
Quantum spin Hall insulator/metal interfaces are formed in graphene ribbons with intrinsic spin-orbit coupling by selectively doping two regions creating a potential step. For a clean graphene ribbon, the transmission of the topological edge states through a n-n or p-p junction is perfect irrespective of the ribbon termination, width, and potential step parameters due to the orthogonality of incoming and outgoing edge channels.  This is shown numerically for an arbitrary crystallographic orientation of the ribbon and proven analytically for zigzag and metallic armchair boundary conditions. In disordered ribbons, the orthogonality between left- and right-movers is in general destroyed and backscattering sets in. However, transmission approaches one by increasing the ribbon's width, even in the presence of strong edge roughness.
\end{abstract}


\maketitle

\newcommand{\e}{\mathrm{e}}
\renewcommand{\i}{\mathrm{i}}
\newcommand{\spinor}[2]{\left(\begin{array}{c} #1 \\ #2 \end{array}\right)}

\section{Introduction}
\begin{figure}
\includegraphics[width=8cm]{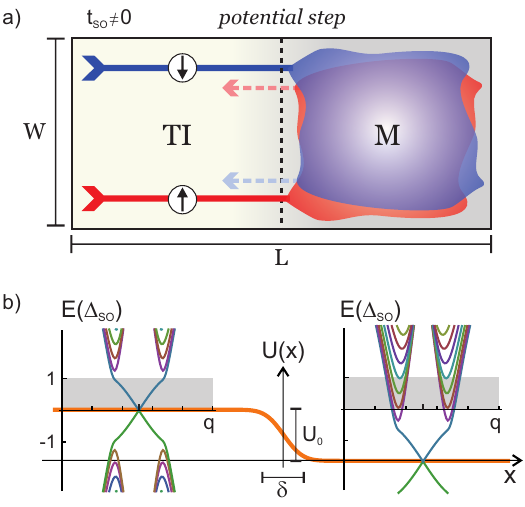}
\caption{\label{Fig:system} (Color online) a) A graphene ribbon with intrinsic spin-orbit coupling is a topological insulator with gapless spin-polarized edge states at low energies. In the presence of a potential step of height $U_0$ bigger than the spin-orbit gap $\Delta_{\rm so}$, a quantum spin Hall insulator/metal interface is formed where topological edge channels evolve into multiple bulk propagating modes. b) Energy bands for a zigzag terminated ribbon at the left and right sides of a potential step of height $U_0$ and width $\delta$. The energy range of interest is marked in grey. 
}
\end{figure}

Since the discovery of the quantum Hall effect~\cite{vonKlitzing1980}, edge states have played an important role in condensed matter physics. With the prediction of the topological insulator phase~\cite{Kane2005, Bernevig2006a, Bernevig2006b} and its observation in \textrm{HgTe/CdTe} quantum wells~\cite{Koenig2007, Roth2010}, the study of edge state physics has received renewed attention, both from the theoretical and experimental side \cite{Knez2011}. This new state of matter is identified by an energy gap in the bulk with gapless states propagating around the edges of the sample. In contrast to the normal quantum Hall effect, these edge states preserve time-reversal symmetry and must be spin-polarized as a consequence: at the system's boundary a pair of counterpropagating spin-up and spin-down states exists as sketched in Fig.~\ref{Fig:system}. As equal-spin states with opposite velocity are thus spatially separated at opposite edges of the sample, and since different-spin states at the same boundary form a Kramers' doublet~\cite{Kane2005,Hasan2010}, edge states are protected against backscattering from time-reversal invariant impurities \cite{Gosalbez2012}. This so-called quantum spin Hall (QSH) state has been first studied theoretically in graphene~\cite{Kane2005}, where it results from the intrinsic spin-orbit (SO) coupling induced by the carbon atoms. In order to detect such dissipation-less edge states, it is crucial to understand the interface created between the two-dimensional QSH system and a metallic phase, since quasiparticle transport through such topological edge channels will be ultimately influenced by the contact resistance of the experimental device \cite{Xia2011}.

Graphene inherits interesting properties from its linear dispersion for low-energy quasi-particle excitations~\cite{Novoselov2005, Zhang2005}, making it a candidate for future spintronics applications~\cite{CastroNeto2009, DasSarma2010}. Part of its functionality comes from the ability to tune the carrier density and carrier type (electron or hole) with electric gates, which allows for the straightforward creation of n-p junctions in graphene. Such junctions allow the investigation of the dynamics of the effective massless Dirac-fermions in graphene, showing interesting properties such as Klein tunneling~\cite{Katsnelson2006, Young2009}, or beam collimation and lensing~\cite{Cheianov2007}. Graphene's functionality can be extended by forming junctions with different materials, such as ferromagnets~\cite{Cho2007, Tombros2007} or superconductors~\cite{Beenakker2006, Heersche2007}. With the advent of topological insulators, the investigation of n-p junctions in the QSH phase (in graphene~\cite{Yamakage2009,Yamakage2010} and in quantum wells\cite{Zhang2010}) as well as interfaces between this phase and other materials \cite{Fu2009, Adroguer2010, Bai2010, Mondal2010, Yokoyama2009, Stanescu2010,Novik2010,Bercioux2010,Mahfouzi2010} has just been initiated.

Unfortunately, the value of the SO-induced gap in graphene is much smaller \cite{Huertas-Hernando2006,Min2006,Yao2007,Boettger2007,Gmitra2009} than initially predicted by Kane and Mele \cite{Kane2005}. Current estimations \cite{Gmitra2009} for pure, flat graphene render a gap of the order of $10\rm \mu eV$ and thus, it is in practice unobservable. However, several efforts to find ways to increase this value can be found recently in the literature. For example, it's been suggested that SO interactions are more important than previously thought in curved graphene \cite{Lopez-Sancho2010,CastroNeto2009}, or can even be induced by means of an external metallic gate \cite{Pereg-Barnea2010}. In particular, Weeks et al. \cite{Weeks2011} proposed a very promising way to engineer a robust QSH state in graphene via adatom deposition, predicting detectable values for the gap of order $10\rm  meV$, while still preserving the effective, graphene-only model of Ref. \onlinecite{Kane2005}. More recently, giant values ($\sim 200\rm meV$) have been calculated for the topological insulator gap in graphene with 5d adatoms \cite{Hu2012}. However, in this last case, the system's physics cannot be described by an effective graphene-only model.

With the perspective that the QSH effect may be achieved in graphene one day \cite{Weeks2011,Hu2012}, in this paper we theoretically study the transmission properties of the QSH edge states in graphene ribbons through an electrostatic potential step, defined for example by selectively doping left and right regions of the ribbon by means of two back gates. When the potential step height $U_0$ is such that the Fermi energy in both regions lies inside the SO induced band gap $\Delta_{\rm so}$, perfect transmission is trivially expected since transport in both regions occurs in the form of topological edge channels. Spin-preserving backscattering processes then amount to crossing the sample's width $W$ and are forbidden as long as $W$ is larger than the width of the edge states~\cite{Strom2009}.
%
However, when $U_0$ is bigger than $\Delta_{\rm so}$, a QSH insulator/metal junction is formed and topological edge channels in one region evolve into multiple bulk channels above the bandgap in the other (see Fig.~\ref{Fig:system}). Backscattering of the QSH edge states could then in principle occur through the metallic region where a path opens up allowing the carriers to cross the sample's width from one boundary to the other. This reflection process would lead to a spin-current parallel to the interface (a similar spin-current has been predicted at an interface between semi-infinite graphene planes in the presence of Rashba SO coupling~\cite{Yamakage2010}). However, we show in this paper that the topological edge states in a clean ribbon of an arbitrary crystallographic orientation cannot be reflected at a QSH insulator/metal junction (either n-n or p-p) due to the orthogonality of left- and right-moving edge states at opposite boundaries. A similar mechanism was recently shown to be responsible for the perfect transmission of the zero Landau level edge state at a magnetic flux step in folded graphene nanoribbons~\cite{Prada2010}. Perfect transmission is also present in n-p junctions, but only when the ribbon is in the armchair or antizigzag configuration, because only then does the n-p junction not introduce intervalley scattering~\cite{Akhmerov2008}. The orthogonality between edge states is proven by deriving analytic expressions for the QSH edge wavefunctions in zigzag and armchair terminated ribbons. For general ribbon terminations, the absence of backscattering is furthermore validated by numerical calculations. Interestingly, we also find that the topological edge state width depends on the ribbon's termination. For an armchair ribbon, this width only depends on the magnitude of the SO induced gap, whereas it is practically independent of this parameter for zigzag ribbons where the edge state width becomes energy-dependent instead. This observation is further confirmed by numerical simulations of the charge density across the sample. These results for clean junctions are analyzed in Section~\ref{Sec:Ballistic} (numerical results) and in Section \ref{Sec:Analytic} (analytical results).

The influence of disorder, produced for example by edge roughness, is analyzed in Section~\ref{Sec:Disorder}. In general, the orthogonality between equal-spin left- and right-movers is then destroyed due to intervalley scattering and the conductance of the junction decreases as inter-boundary backscattering becomes possible through the metallic region. Backscattering is also present for intrinsic sources of intervalley scattering, like a (clean) n-p junction in a zigzag ribbon~\cite{Akhmerov2008}. However, we find that in any case, perfect transmission is restored upon increasing the ribbon's width $W$, essentially because the probability of forward scattering increases with $W$ while the only available backscattering channel remains one. The conductance recovery with increasing $W$ depends on the ribbon termination due to the different edge state character for different ribbon boundaries. A summary of all our results and some concluding remarks are given in Section~\ref{Sec:Conclusions}.

\section{Transport in a ballistic junction}\label{Sec:Ballistic}

\subsection{System description}
We consider the system depicted schematically in Fig.~\ref{Fig:system}. It consists of a graphene ribbon subject to intrinsic SO coupling which opens a gap in the bulk graphene spectrum. The Fermi energy in the left part of the ribbon is considered to lay inside this gap, so that transport in this region can only take place through the topological edge channels (see Fig.~\ref{Fig:system}). In the right part of the ribbon, the potential is shifted upwards or downwards by an electrostatic gate in order to reach doping levels above the SO induced bulk gap. Topological edge states then do not play a role anymore in this region and transport is mediated by normal propagating bulk modes. In this way we create a QSH insulator/metal junction in graphene (although the SO interaction is present throughout all the ribbon). The Hamiltonian of our system is given by
\begin{equation}
H = H_0 + H_{\rm so} + H_{\rm U}.\label{TotalH}
\end{equation}
The first term represents the standard hopping between nearest-neighbor carbon atoms,
\begin{equation}
H_0 = -t \sum_{\langle ij \rangle, s_z} c_{i,s_z}^{\dag} c_{j,s_z},\label{H0}
\end{equation}
where $c_{i,s_z}^{\dag}$ is the creation operator for an electron with spin $s_z$ on site $i$. The second term in the Hamiltonian accounts for the intrinsic SO coupling introduced by Kane and Mele~\cite{Kane2005}:
\begin{equation}
H_{\rm so} = \mathrm{i} t_{\rm so} \sum_{\langle\langle ij \rangle \rangle, s_z} \nu_{ij} \, s_z\,c_{i,s_z}^{\dag} c_{j,s_z},\label{Hso}
\end{equation}
where the summation runs over next-nearest neighbors (note that in Eq. (\ref{H0}) hopping occurs between different sublattices $A$ and $B$, while in (\ref{Hso}) hopping is within the same sublattice). The prefactor $\nu_{ij} = +1(-1)$ if the electron takes a left (right) turn to reach its next-nearest neighbor. Since this term commutes with the electron's spin $s_z$, the Hamiltonian of the system decouples into two independent Hamiltonians for spin-up and spin-down that can be considered independently. In this case, we have edge states for each boundary with spin-up moving in one direction and spin-down moving in the opposite one (see Fig. \ref{Fig:system}). In the presence of spin-mixing terms that do not break time-reversal symmetry (like for example the Rashba term), conserved currents do not have a well-defined spin projection. Nevertheless, it will still be possible to define two sets of counter-propagating currents at each boundary that are topologically protected against non-magnetic impurities~\cite{Kane2005,Hasan2010}. For simplicity, below we consider the case where $s_z$ is a good quantum number, but the physics we describe should be analogous if time-reversal symmetric spin-mixing terms were present (as long as they are constant in space and the magnitude of their coupling smaller than the intrinsic SO one). 

The last term $H_{\rm U}$ in the Hamiltonian describes the on-site graphene doping, given by
\begin{equation}
H_{\rm U} = \sum_{i,s_z} U(x) c_{i,s_z}^{\dag} c_{i,s_z}.
\end{equation}
The sketch in Fig. \ref{Fig:system} shows a n-n junction. In this case the doping profile along the boundary direction can be modeled by a smooth potential step of height $U_0$ of the form
\begin{equation}
U(x) = -\frac{U_0}{2} \left[ 1 + {\rm erf} \left( \frac{2(x-L/2)}{\delta}\right) \right],\label{U(x)}
\end{equation}
where $L$ is the length of the ribbon and $\delta$ controls the smoothness of the step. This gives $U(x) = 0$ when $x-L/2 <\delta$ (left part of the sample), while $U(x) = - U_0$ for $x-L/2 >\delta $ (right part of the sample). In order to have a QSH insulator/metal junction, we need $U_0>\Delta_{\rm so}$ and consider the energy range $[0, \Delta_{\rm so}]$ depicted in grey in Fig. \ref{Fig:system}(b). Here $\Delta_{\rm so}$ is the SO induced bulk gap, given by $\Delta_{\rm so}\equiv 3 \sqrt{3} t_{\rm so}$. An equivalent p-p junction is obtained by changing $U_0$ by $-U_0$ in Eq.~(\ref{U(x)}), and considering energies between $-\Delta_{\rm so}$ and $0$. Furthermore, we could consider n-p (or p-n) junction if we chose $U_0>2\Delta_{\rm so}$. In the following we will focus on the n-n junction shown in Fig.~\ref{Fig:system} for simplicity.
%
\begin{figure} 
\includegraphics[width=8cm]{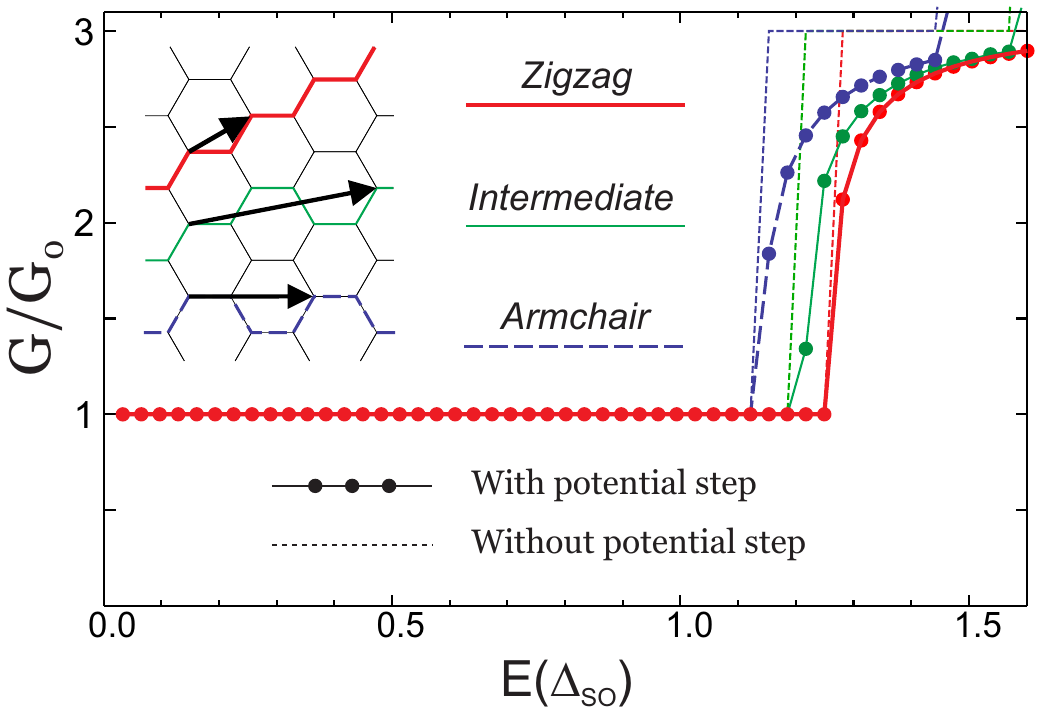}
\caption{\label{Fig:Tclean}(Color online)  Conductance vs Fermi energy for a QSH insulator/metal n-n junction in the ballistic regime (see Fig. \ref{Fig:system}). Perfect transmission is found irrespective of the ribbon termination: zigzag (solid red line), armchair (dashed blue line), and an intermediate crystallographic orientation (thin green line). All ribbons have a width $W = 5.6\lambda_{\rm so}$. The potential step height is $U_0 = 1.6\Delta_{\rm so}$ and its width $\delta = 0.5\lambda_{\rm so}$.  
}
\end{figure}

\subsection{Perfect transmission}
At a QSH insulator/metal junction in a graphene ribbon, an incoming edge state may in principle either be transmitted or reflected at the junction. Spin-preserving
reflection is only possible when the electron is transferred from the incoming edge state into the counterpropagating edge state at the opposite boundary. Such an inter-boundary reflection is in principle possible because the electron's wavefunction to the right of the junction is not constrained to the edges, but spans the whole bulk in multiple propagating metallic channels. However, one of the main points of our paper is that, at a valley-preserving junction, this reflection process is forbidden due to the orthogonality of forward and backward topological edge channels, much like in the Klein paradox. In Fig.~\ref{Fig:Tclean}, the conductance of the junction in units of $G_0=2e^2/h$ is plotted as a function of the Fermi energy of the incoming electrons (in units of $\Delta_{\rm so}$). In all our simulations we have used $t_{\rm so} = 0.03 t$, where $t$ is the nearest-neighbor hopping amplitude. This is the unrealistically large value initially given in Ref. \onlinecite{Kane2005}. We use this value in order to be able to perform our numerical simulations since, for realistic carbon-only induced SO coupling \footnote{Ref. \onlinecite{Gmitra2009} gives an estimate of the SO-induced gap of $\Delta_{\rm so}=24\mu \rm eV$, which corresponds to $t_{\rm so} \approx 2\times 10^{-6} t$. In the proposal of engineered graphene via adatom deposition of Ref. \onlinecite{Weeks2011}, $\Delta_{\rm so}\approx 10 \rm meV$, which corresponds to $t_{\rm so} \approx 7\times 10^{-4} t$. Finally, Ref. \onlinecite{Hu2012} finds a giant value $\Delta_{\rm so}\approx 200\rm meV$, for which $t_{\rm so} \approx 0.014 t$, similar to the one used in our numerics. However, and as mentioned by the authors, in their proposal the system cannot be described by an effective graphene-only model anymore.}, we would need to simulate too wide ribbons, which is beyond our computational capabilities. However, and using scaling arguments, it is possible to argue that a system with smaller $\tilde{t}_{\rm so}=\alpha t_{\rm so}$ ($\alpha\ll 1$) would exhibit the same transport behaviour as long as the width of the ribbon is also scaled by $\tilde{W}=W/\alpha$, and the Fermi energy by $\tilde{E}=\alpha E$. \cite{Prada2010c}

In Fig. \ref{Fig:Tclean} different ribbon terminations are considered: armchair (parametrized by an angle $\theta=0$, dashed blue curve), zigzag ($\theta=\pi/6$, solid red curve), and an intermediate crystallographic orientation ($\theta=0.4\pi/6$, thin green curve). The ribbon's width is $W=54a\approx 5.6 \lambda_{so}$, where $a$ is the interatomic distance and $\lambda_{so}\equiv\hbar v_F/\Delta_{so} \approx 10 a$ is the SO length. Here $v_F=3a t/2\hbar$ is the quasiparticle Fermi velocity. The potential step has parameters $U=0.25t\approx 1.6\Delta_{\rm so}$ and $\delta=5a\approx 0.5\lambda_{so}$. Perfect transmission is manifested by a flat first conductance plateau at energies where only edge states are present on the QSH insulator side. Note that the energy window of such plateau, i.e., the effective gap, is slightly bigger than $\Delta_{\rm so}\equiv 3 \sqrt{3} t_{\rm so}$. This is due to confinement effects in finite width ribbons. When $W\rightarrow\infty$, the effective gap tends to $\Delta_{\rm so}$. For higher energies, transport is carried by additional propagating modes above the SO gap that are not protected by orthogonality and thus the transmission is reduced (compared to the situation without a potential step, dotted thin lines). We have checked numerically that perfect transmission below the gap holds irrespective of the ribbon's crystallographic orientation and its width $W$,\footnote{For semiconducting armchair ribbons, $W$ has to be much bigger than the edge state width $\lambda$ so that the ribbon in the presence of intrinsic SO coupling is metallic.} and is independent of $U_0>0$ and $\delta$ (as long as $\delta$ is larger than a few carbon-carbon lattice constants). As mentioned above, perfect transmission also holds irrespective of the value of $t_{\rm so}$, although, of course, the smaller $t_{\rm so}$, the smaller the induced SO gap and, therefore, the smaller the energy window where this effect is observable. 

\begin{figure} 
\includegraphics[width=8cm]{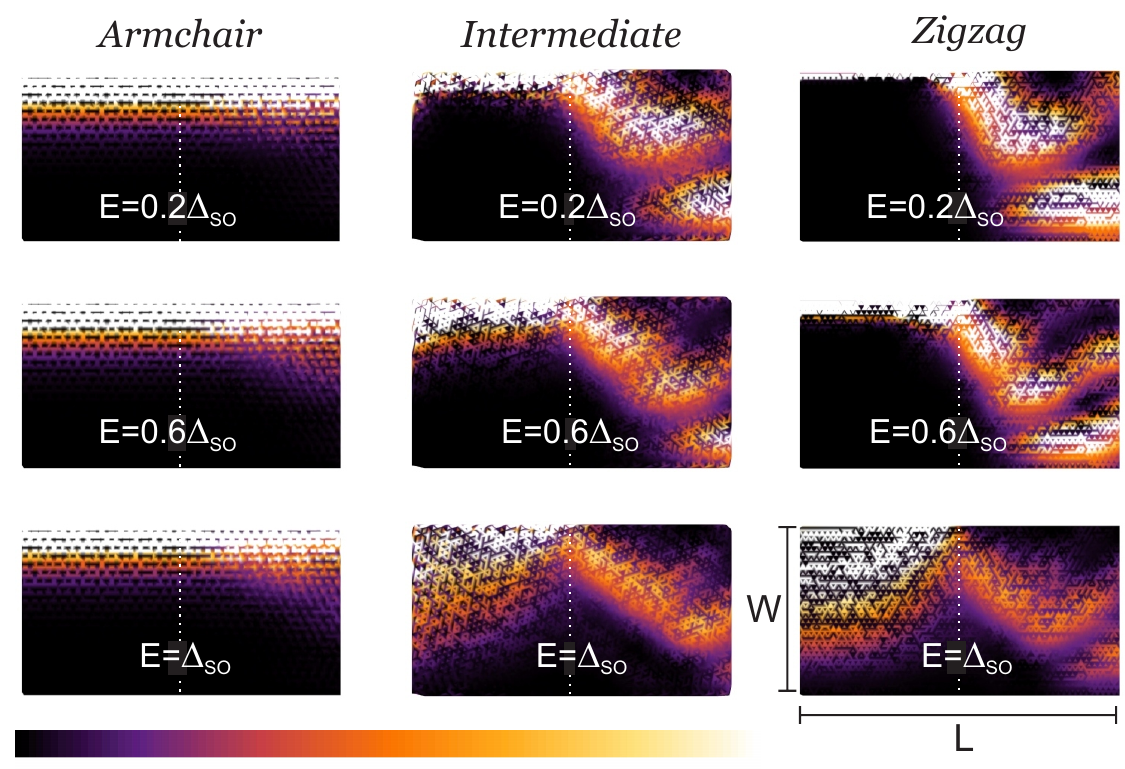}
\caption{\label{Fig:chargedensity}(Color online) Numerical results for the charge density across a potential step ($U_0=1.6\Delta_{\rm so}$ and $\delta=0.5\lambda_{\rm so}$) for a clean ribbon of $W=5.6\lambda_{\rm so}$. A section $L=10.4\lambda_{\rm so}$ of the infinitely long ribbon is shown, and various ribbon terminations and Fermi energies are considered. At the left of the n-n junction (placed in the middle of each flake, vertical dotted white line), transport is carried by a topological edge channel that evolves into multiple normal channels at the right. A similar contribution with opposite spin comes in from the lower boundary (not shown). 
}
\end{figure}

Our numerical results for mesoscopic transport are obtained using the recursive Green's function technique~\cite{Datta1997}. This technique is suitable for the computation of quasiparticle transport through systems that are described with a tight-binding Hamiltonian and composed typically of a central part of finite dimensions (where all the scattering happens) joined to two semi-infinite leads. To calculate the scattering matrix of the system, one computes by recursive decimation of the central region the Green's functions that describe the propagation between leads. The contact self-energy coming from the leads, that must be added to the central system sites connected to the leads, is moreover computed by solving an equivalent generalized eigenvalue problem numerically. An extension of the method using double-sweep decimation \cite{Metalidis2005} gives us also access to the local charge density in real space. This allows for a direct visualization of the shape of the edge states and their propagation across the potential step, as depicted in Fig.~\ref{Fig:chargedensity} for different ribbon terminations and energies below the bulk effective gap. The charge density is shown in arbitrary units, with black representing the absence of charge, and white its maximum value. Since spin-up and spin-down are decoupled in our problem, they can be analyzed independently and their contribution summed up at the end. In Fig.~\ref{Fig:chargedensity} only the spin current incoming from the upper boundary is considered. A similar contribution (mirror symmetric with respect to the ribbon axes) coming in from the lower boundary with the opposite spin is not shown. In all ribbons, the incoming topological edge state in the left region propagates towards the junction and then spreads out into multiple channels available in the bulk of the metallic region. There is no sign of a backscattered edge state at the lower boundary on the QSH insulator side. It is interesting to note that the width of the topological edge state clearly depends on the crystallographic orientation of the ribbon: it is energy-independent for an armchair ribbon while it strongly depends on energy for the zigzag and intermediate one. This behavior is analytically confirmed in the next Section and will have consequences in the presence of disorder, as analyzed in Section~\ref{Sec:Disorder}. 

\section{Analytical results}\label{Sec:Analytic}

For armchair and zigzag graphene ribbons we were able to derive analytic expressions for the topological edge state wavefunctions. For clean metallic ribbons, this allows us to prove the orthogonality of forward and backward channels at opposite boundaries that is at the basis of the perfect transmission observed in Fig.~\ref{Fig:Tclean}. This orthogonality is in the valley sector of the wavefunction, which we assume must be preserved in the presence of a smooth electrostatic potential step. Furthermore, our analytical results also offer an explanation for the different edge state widths for armchair and zigzag ribbons observed in Fig.~\ref{Fig:chargedensity}.

\begin{figure}
	\includegraphics[width = 0.7\columnwidth]{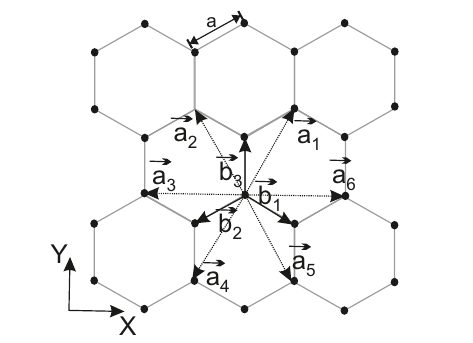}
	\caption{\label{Fig:graphenegeometry} Geometry of the graphene lattice considered in the derivation of the edge state wavefunctions. Here $\vec{b}_{1,2}=a(\pm\sqrt{3}/2,-1/2)$, $\vec{b}_3=a(0,1)$, $\vec{a}_{1,2}=a(\pm\sqrt{3}/2,3/2)$, $\vec{a}_{6,3}=a(\pm\sqrt{3},0)$ and $\vec{a}_{5,4}=a(\pm\sqrt{3}/2,-3/2)$, where $a$ is the interatomic distance.}
\end{figure}

We start by considering the lattice and the reference system shown in Fig.~\ref{Fig:graphenegeometry}. In terms of the lattice vectors $\vec{b}_i$ (linking nearest-neighbor carbon atoms) and $\vec{a}_i$ (linking next-nearest neighbors), the tight-binding Hamiltonians of Eqs. (\ref{H0}) and (\ref{Hso}) can be rewritten as
\begin{equation}
H_0=-t \sum_{i,s_z}\sum_{n=1}^3 a^{\dagger}_{\vec{r}_i,s_z} \, b_{\vec{r}_i+\vec{b}_n,s_z}+\rm{h.c.},
\end{equation}
and
\begin{eqnarray}
H_{\rm so}= \i t_{\rm so}\sum_{i,s_z} \sum_{n=1}^6 (-1)^{n+1} s_z\left[a^{\dagger}_{\vec{r}_i,s_z} \, a_{\vec{r}_i+\vec{a}_n,s_z}\right.\nonumber\\- b^{\dagger}_{\vec{r}_i+\vec{b}_3,s_z} \,b_{\vec{r}_i+\vec{b}_3+\vec{a}_n,s_z} \left.\right].
\end{eqnarray}
In the previous two expressions, $i$ runs over unit cells, $a^{\dagger}$ and $b^{\dagger}$ are creation operators on sublattice $A$ and $B$ respectively, and $s_z =\pm 1$ is the z-component of the electron spin.

Fourier transforming these Hamiltonians into momentum space (with a suitable phase origin for each sublattice), one gets:
\begin{equation}
H_0=\sum_{\vec{q},s_z} \xi(\vec{q}) \, a^{\dagger}_{\vec{q},s_z} \, b_{\vec{q},s_z} + \textrm{h.c.},
\end{equation}
where $\vec{q}$ is the wave vector in the graphene plane measured from the center of the first Brillouin zone and
\begin{equation} 
\xi(\vec{q}) =-t \sum_{n=1}^3 \e^{-\i \vec{q} \cdot \vec{b}_n} = -t\left[2\cos(\frac{\sqrt{3}}{2}a q_x)+\e^{-i\frac{3}{2}a q_y}\right]. \label{eq:defxi}
\end{equation}
On the other hand
\begin{equation}
H_{\rm so}=\sum_{\vec{q},s_z} \chi(s_z,\vec{q}) \, \left(a^{\dagger}_{\vec{q},s_z} a_{\vec{q},s_z}-b^{\dagger}_{\vec{q},s_z} b_{\vec{q},s_z}\right),
\end{equation}
with
\begin{eqnarray} \label{eq:defchi}
&&\chi(s_z, \vec{q})= \i s_z t_{\rm so} \sum_{n=1,...,6} (-1)^{n+1} \e^{-\i\vec{q} \cdot \vec{a}_n}\\
&=&2 s_z t_{\rm so}\left[2\sin(\frac{\sqrt{3}}{2}a q_x)\cos(\frac{3}{2}a q_y)-\sin(\sqrt{3}a q_x)\right]. \nonumber
\end{eqnarray}
The total Hamiltonian can thus be written as
\begin{equation} \label{eq:totalhamiltonian}
H=\sum_{\vec{q},s_z}(a^{\dagger}_{\vec{q},s_z}\;b^{\dagger}_{\vec{q},s_z})\left(\begin{array}{cc}
\chi(s_z,\vec{q}) & \xi(\vec{q}) \\
\xi^*(\vec{q}) & -\chi(s_z,\vec{q})\end{array} \right)
\left(\begin{array}{c}
a_{\vec{q},s_z} \\ b_{\vec{q},s_z}\end{array} \right),
\end{equation}
with eigenvalues
\begin{equation} \label{eq:generaldispersion}
E(\vec{q})=\pm\sqrt{|\xi(\vec{q})|^2+\chi(s_z,\vec{q})^2}
\end{equation}
and eigenvectors
\begin{equation} \label{eq:spinors}
	\spinor{\psi_A(s_z,\vec{q})}{\psi_B(s_z,\vec{q})} \propto \spinor{\xi(\vec{q})}{E - \chi(s_z, \vec{q})}.
\end{equation}
The total wave function is then
\begin{equation}
\Psi_{s_z,\vec{q}}(x,y) = \e^{\i q_x x} \e^{\i q_y y}\left(\begin{array}{c}
\psi_A(s_z, \vec{q}) \\ \psi_B(s_z,\vec{q})\end{array} \right).\label{eq:totalwf}
\end{equation}

\subsection{Edge states at an armchair boundary}

First we study the easier case, the armchair ribbon. We consider it extends along the horizontal $x$-direction, so that the boundaries are at $y=0$ and $y=W$ (thus the lattice is simply rotated $90^{\circ}$ with respect to Fig. \ref{Fig:graphenegeometry}). For simplicity, we assume that the ribbon is wide enough so that the two opposite boundaries can be treated independently.
 
In the armchair case, it is known \cite{Prada2010b} that the topological edge states have wavefunction solutions around the two inequivalent $K_1$ and $K_2$ points of the first Brillouin zone (with $\vec{K}_{1,2}=\pm \frac{4\pi}{3\sqrt{3}a}\hat{y}$).  This means that, for small energies, we can expand the Hamiltonian (\ref{eq:totalhamiltonian}) to first order in momentum around those points writing $\vec{q}=\vec{K}_{1,2}+\vec{k}$, where $\vec{k}=(k_x,k_y)$ is the in-plane wave vector measured from the $K_{1,2}$ point. This gives $\xi\approx \hbar v_F(\tau_z k_y+i k_x)$ and $\chi\approx -s_z\tau_z\Delta_{\rm so}$, where $\tau_z=\pm 1$ specifies the Dirac point $K_1$ and $K_2$, and it's known as the valley degree of freedom. The Hamiltonian matrix in Eq. (\ref{eq:totalhamiltonian}) can then be written as
\begin{equation}
\mathcal{H}_{\tau_z,s_z}(\vec{k})=\hbar v_F (k_y\sigma_x\tau_z-k_x\sigma_y)-\Delta_{\rm so}\sigma_z\tau_z s_z,\label{HDirac}
\end{equation}
where $\sigma_i$ are Pauli matrices representing the pseudospin degree of freedom (corresponding to the two sites per unit cell of the graphene lattice). The eigenenergy for this Hamiltonian is $E = \pm \sqrt{\Delta_{\rm so}^2 + (\hbar v_F)^2(k_x^2 + k_y^2)}$, and the pseudospin spinor is
\begin{equation}
\spinor{\psi^{\tau_z}_{A}(s_z,\vec{k})}{\psi^{\tau_z}_{B}(s_z,\vec{k})} \propto\spinor{\hbar v_F  (\tau_z k_y +\i k_x)}{E+ s_z\tau_z \Delta_{\rm so}}.
\end{equation}
As we are looking for an edge state solution, the wavefunction has to decay from the edge into the bulk. Considering the lower boundary (the semi-infinite graphene sheet extends towards $y>0$), we then substitute $k_y\rightarrow \i \kappa$ in the previous expressions. For armchair ribbons we know that the boundary condition amounts to setting the wavefunction to zero at $y=0$, both on $A$ and $B$ sublattice sites \cite{Brey2006}. To obtain a nontrivial solution, we have to consider a superposition of two wavefunctions. Here one can choose a combination of waves around the $K_1$ and $K_2$ points, because the low-energy limit corresponds to having wavevectors in the vicinity of these points. Our ansatz is thus
\begin{equation} \label{eq:armchairwvf}
\Psi_{\rm ac} (\vec{x}) = \e^{\i k_x x} \e^{-\kappa y} \left[ \alpha_0 \e^{\i Ky}\! \spinor{\psi_{A}^{+}}{\psi_B^+} + \beta_0 \e^{-\i Ky}\!\spinor{\psi_A^-}{\psi_B^-} \right],
\end{equation}
where $\alpha_0, \beta_0$ are the relative amplitudes of the wavefunction in each valley (and $K=|\vec{K}_{1,2}|$). A nontrivial wavefunction~(\ref{eq:armchairwvf}) fulfilling the boundary condition can only be found when
\begin{equation}
	\left|
	\begin{array}{cc}
	\psi_A^+ & \psi_A^- \\
	\psi_B^+ & \psi_B^-
	\end{array}
	\right|=0.
\end{equation}
This imposes the condition $\kappa = s_z k_x \Delta_{\rm so}/E$. Inserting it into the eigenenergy one obtains
\begin{equation}
E = \pm \hbar v_F |k_x|,
\end{equation} 
where the $\pm$ signs correspond to quasiparticles in the conduction and the valence band. This means that the dispersion of these topological states is linear at small energies. Moreover, it is independent of the SO strength for the armchair case. The result for $\kappa$ can then be written as: $\kappa = (\Delta_{\rm so}/\hbar v_F) \, s_z \, \textrm{sgn}(k_x) \, \textrm{sgn}(E)$.
%
%

At the lower boundary, $\kappa$ should be positive to get a decaying solution, so, e.g., electrons ($E > 0$) propagate either with spin-up to the right ($k_x > 0$), or with spin-down to the left $k_x <0$. The pseudospin associated with these electrons is
\begin{equation}
	\spinor{\psi_A^{\tau_z}}{\psi_B^{\tau_z}}^l=\frac{1}{\sqrt{2}}\spinor{\i s_z}{1}, \label{Pspinl}
\end{equation}
independent of the valley. At the upper boundary one needs $\kappa < 0$ and the spins are reversed for the same directions. The pseudospin is then, again for $E>0$,
\begin{equation}
	\spinor{\psi_A^{\tau_z}}{\psi_B^{\tau_z}}^u=\frac{1}{\sqrt{2}}\spinor{-\i s_z}{1}.
\end{equation}

The width of the edge state is given by $\lambda = 1/|\kappa|$, and is thus determined solely by the SO coupling strength. For a wide armchair ribbon we have then
\begin{equation}\label{lambdaAC}
\lambda^{\rm ac}=\hbar v_F/\Delta_{\rm so}\equiv\lambda_{\rm so}.
\end{equation}
In our numerical calculations with $t_{\rm so} = 0.03 t$ we get a decay length ${\lambda_{\rm ac}} \approx 10 a$. This is in good agreement with the edge state width observed in Fig.~\ref{Fig:chargedensity} for the armchair case.  



In order to prove the orthogonality between equal-spin incoming and outgoing edge channels, it is convenient to express the previous Hamiltonian (\ref{HDirac}) in the \textit{valley isotropic basis},\cite{Akhmerov2007} where the transformed Hamiltonian is the same for both valleys. Leaving $\mathcal{H}_{\tau_z,s_z}$ unchanged for $\tau_z=+1$, we transform the one for the other valley, $\tau_z=-1$, such that $\mathcal{H}^{\rm iso}_{-1,s_z}=\mathcal{H}_{+1,s_z}$. This is done with the unitary transformation $\mathcal{H}^{\rm iso}_{-1,s_z}=U\mathcal{H}_{-1,s_z}U^{+}$, where $U=i \sigma_y$. The valley isotropic Hamiltonian matrix is then
\begin{equation}
\mathcal{H}^{\rm iso}_{\tau_z,s_z}(\vec{k})=\hbar v_F (k_y\sigma_x-k_x\sigma_y)-\Delta_{\rm so}\sigma_zs_z.\label{HDiracIso}
\end{equation}
In a similar way, the psedospin spinor for $\tau_z=-1$ is also changed: $\psi^{\rm iso}_{-}=U\psi_{-}$. The total wave function for the topological edge channel in this basis is
\begin{equation} \label{eq:armchairwvfiso}
\Psi^{\rm iso}_{\rm ac} (\vec{x}) = \e^{\i k_x x} \e^{-\kappa y} \left[ \alpha \e^{\i Ky}\! \spinor{\psi_{A}^{+}}{\psi_B^+} + \beta \e^{-\i Ky}\!\spinor{\!\psi_B^-}{\!-\psi_A^-} \right],
\end{equation}
where now $\alpha$ and $\beta$ represent the valley polaritazion or isospin of the wave function in the valley isotropic basis, that can be represented by the spinor $|\tau\rangle\equiv (\alpha,\beta)$.

A potential step conserves the isospin as long as it is smooth on the scale of the lattice constant.
It is possible to see that the isospin for the lower boundary satisfying the boundary conditions [and using Eq. (\ref{Pspinl})] is given by $|\tau\rangle^l=\frac{1}{\sqrt{2}}(\i s_z,1)$. Equivalently, for the upper boundary we have $|\tau\rangle^u=\frac{1}{\sqrt{2}}(-\i s_z e^{-2\i KW},1)$. Writing $W = \frac{\sqrt{3}}{2}(N+1)a$ in terms of the number $N$ of carbon atoms along the ribbon's width, \footnote{Note that the actual width of the ribbon is $W'=\frac{\sqrt 3}{2}(N-1)a$, but the wave function has to vanish one site away from the ribbon's boundary at each side of the ribbon, so that $W=W'+\sqrt 3 a$.} one can see that the isospins of counterpropagating equal-spin edge states (moving at opposite boundaries) are orthogonal when the armchair ribbon is metallic, i.e., for $N=3n-1$, with $n=1,2,...$. Scattering between these states is thus forbidden, which leads to the absence of reflection. 

The semiconducting armchair ribbon is a special case. Let us show this with an example. Imagine we had a ribbon without potential step but with a smooth intrinsic SO step instead. The left region is then exactly the same as in Fig. \ref{Fig:system}, but the right region now is plain graphene (i.e., without SO coupling). For metallic armchair ribbons (actually for zigzag and intermediate ones too), such a junction constitutes a QSH insulator/metal interface and we observe numerically a perfect transmission for energies below the induced SO bulk gap. However, the semiconducting armchair ribbon is insulating for small energies in the right region, and the transmission is zero for such energies. Therefore, knowledge of the wavefunction in the left region only is not sufficient in order to make a statement for the transmission in semiconducting armchair ribbons, due to its intrinsic bandgap. In our case of a constant SO coupling and a potential step, we can only state that, numerically, we also observe perfect transmission as long as the ribbon is metallic in the presence of SO coupling, i.e., $W\gtrsim 2 \lambda_{\rm so}$, even though the isospins of the incoming and outgoing edge states are not orthogonal.

\subsection{Edge states at a zigzag boundary}\label{Sec:Ballistic-zigzag}
Now we turn to a zigzag ribbon. Like before, we consider that it extends along the horizontal $x$-direction, with boundaries at $y=0$ and $y=W$ (Fig. \ref{Fig:graphenegeometry}). Again, we assume that the ribbon is wide enough so that the two opposite boundaries can be treated independently. In contrast to the armchair case, for zigzag ribbons the dispersive topological edge states are centered around the $M$ point of the first Brillouin zone (see below). Therefore, the standard low energy expansion of the Hamiltonian around the $K_{1,2}$ points (that leads to the Dirac equation in the absence of SO interaction) cannot be made, since the low-energy solutions in the presence of the intrinsic SO coupling occur for momenta far away from those points. Thus, we have to start with the full tight-binding Hamiltonian to calculate the wavefunctions and dispersion of the topological edge states \cite{Zarea2009}.

In the absence of SO coupling it is known that the boundary condition for zigzag ribbons amounts to a vanishing wavefunction on one of the sublattices \cite{Brey2006}. However, as SO coupling induces next-nearest neighbor hopping, the total wave function in our case has to vanish in both the $A$ and the $B$ sublattice sites at the edge. This condition can never be met by a single wavefunction of the form~(\ref{eq:totalwf}), so we will look at a superposition of two of them. Furthermore, to obtain an edge state localized in the $y$-direction, we replace $q_y \rightarrow \i \kappa$ and make the following ansatz:
%

\begin{eqnarray} \label{eq:ansatz_zigzag}
\Psi_{\rm zz}(\vec{x}) &=& \alpha\, \e^{\i q_x x}\, e^{-\kappa_1 y} \left(\begin{array}{c}
\psi_A(s_z,q_x,\kappa_1) \\ \psi_B(s_z,q_x,\kappa_1) \end{array} \right)\nonumber \\ &+& \beta\, \e^{\i q_x x}\,\e^{-\kappa_2 y} \left( \begin{array}{c}
\psi_A(s_z,q_x,\kappa_2) \\ \psi_B(s_z,q_x,\kappa_2)\end{array}\right).
\end{eqnarray} 

In the following, the notation
\begin{subequations}
\begin{eqnarray}
	Q_c &\equiv& 2 \cos(\frac{\sqrt{3}}{2} q_x a), \\
	Q_1 &\equiv& 2 \sin(\frac{\sqrt{3}}{2} q_x a), \\
	Q_2 &\equiv& 2 \sin(\sqrt{3} q_x a), \\
	\phi_i &\equiv& \e^{-\frac{3}{2} a \kappa_i},  \label{eq:phi_asfuncof_lambda}
\end{eqnarray}
\end{subequations}
will prove to be convenient. The terms $\xi$ and $\chi$ in Eqs.~(\ref{eq:defxi}) and~(\ref{eq:defchi}) can then be written as
\begin{subequations} \label{eq:xichivsphi}
\begin{eqnarray} 
\xi(q_x, \phi_i) &=& -t \left( Q_c + \phi_i^{-1} \right), \\
\chi(s_z, q_x, \phi_i) &=& -t_{\rm so} s_z \left[ Q_2 - Q_1 (\phi_i + \phi_i^{-1})\right],
\end{eqnarray}
\end{subequations}
and the energy dispersion~(\ref{eq:generaldispersion}) becomes
\begin{equation} \label{eq:epsdispersion}
	\epsilon^2 = \left( Q_c + \phi_i^{-1}\right) \left( Q_c + \phi_i \right) + \tau_{\rm so}^2 \left( Q_2 - Q_1 (\phi_i + \phi_i^{-1})\right)^2.
\end{equation}
Here we defined $\epsilon \equiv E/t$, and $\tau_{\rm so} \equiv t_{\rm so}/t$. Multiplying this equation by $\phi_i^2$ gives a fourth order equation in $\phi_i$ from which one can in principle calculate the allowed values of $\phi_i$ as a function of energy. Solving this equation exactly proves to be too difficult. However, since we are interested in small values of the normalised energy $\epsilon$ and SO coupling strength $\tau_{\rm so}$, we can make an expansion to second order in these parameters. From the four solutions of Eq.~(\ref{eq:epsdispersion}), only two fulfill the condition $|\phi_i| < 1$ for a decaying wavefunction around the M-point ($q_x a = \pi / \sqrt{3}$ corresponding to zero energy $\epsilon=0$). They are
\begin{subequations} \label{eq:phivalues}
\begin{eqnarray} 
	\phi_1 &=& -\tau_{\rm so}^2 \frac{Q_1^2}{Q_c}, \\
	\phi_2 &=& -Q_c + \epsilon^2 \frac{Q_c}{Q_c^2-1} \nonumber \\ 
	&\,&\quad - \tau_{\rm so}^2 \frac{(Q_c(Q_2 + Q_c Q_1) + Q_1)^2}{Q_c (Q_c^2-1)}.
\end{eqnarray}
\end{subequations}

The quantities $\phi_i$ still depend explicitly on both energy and wavevector $q_x$. In order to find the dispersion relation $\epsilon(q_x)$, we have to apply the boundary condition $\Psi_{\rm zz}(x,y=0) = \spinor{0}{0}$ to the wavefunction defined in Eq.~(\ref{eq:ansatz_zigzag}). A non-trivial solution for the coefficients $\alpha$ and $\beta$ is then obtained when
\begin{equation}\label{eq:condition}
\left|\begin{array}{cc}
\psi_A(s_z, q_x, \phi_1) & \psi_A(s_z, q_x, \phi_2) \\
\psi_B(s_z, q_x,\phi_1) & \psi_B(s_z, q_x, \phi_2)\end{array}\right|=0.
\end{equation}
%
With Eqs. (\ref{eq:spinors},\ref{eq:xichivsphi}) for the spinors, and inserting the allowed values for $\phi_i$ from Eq.~(\ref{eq:phivalues}), one obtains the dispersion relation (up to first order in $\tau_{\rm so}$):
\begin{equation}
	E(s_z, q_x) = -3t \, s_z \tau_{\rm so} \, Q_2 = -6 \, s_z t_{\rm so} \sin(\sqrt{3} q_x a).
\end{equation}
The spin-up and spin-down bands thus cross at the wave vector $q_x a = \pi/\sqrt{3}$ with zero energy, the $M$ point of the first Brillouin zone (at $q_x a = 0$, the $K_{1(2)}$ point, there is no edge state solution with $|\phi_i| < 1$). Quasiparticles with spin-up (spin-down) thus move with negative (positive) velocity $\frac{1}{\hbar}\frac{\partial \epsilon}{\partial q_x}$ at the lower boundary. At the upper boundary, the role of spin is interchanged.

The $\phi_i$ in Eq.~(\ref{eq:phivalues}) can assume negative values. With the definition $\phi_i \equiv \e^{-\frac{3}{2} a \kappa_i}$, this means that the solutions for $\kappa_i$ can be complex and the wavefunction oscillates in the $y$-direction with a decaying envelope. In particular, we find that the wavefunction changes sign when propagating from one zigzag dimer line to the next. The edge state width has thus to be defined as $\lambda_i\equiv 1/\textrm{Re}(\kappa_i)$. With Eq.~(\ref{eq:phi_asfuncof_lambda}), one obtains
\begin{equation}
	\lambda^{\rm zz}_{1,2} = -\frac{3a}{2 \log(|\phi_{1,2}|)}.\label{lambdaZZ} 
\end{equation}

Until now, we considered the lower boundary. For the upper boundary, the semi-infinite graphene sheet extends now to negative $y$-values. A localized state is then found by substituting $q_y \rightarrow -\i \kappa$ in Eq.~(\ref{eq:totalwf}). This will interchange the off-diagonal terms $\xi$ and $\xi^\star$ in the Hamiltonian~(\ref{eq:totalhamiltonian}). If we further substitute $s_z \rightarrow -s_z$, then the diagonal terms are interchanged too. The Hamiltonian in Eq.~(\ref{eq:totalhamiltonian}) for the upper boundary then looks exactly as the one for the lower boundary, but with the $A$- and $B$-sites interchanged. Using Eq. (\ref{eq:spinors})  and the condition (\ref{eq:condition}), the pseudospin for the lower boundary is
\begin{equation}
	\spinor{\psi_A^{\phi_i}}{\psi_B^{\phi_i}}^l= \spinor{1}{s_z\tau_{\rm so} Q_1}
\end{equation}
to first order in $\tau_{\rm so}$, both for $i=1,2$. With the substitution $A \leftrightarrow B$ and $s_z \rightarrow -s_z$ mentioned above, the pseudospin for the upper boundary is
\begin{equation} 
	\spinor{\psi_A^{\phi_i}}{\psi_B^{\phi_i}}^u = \spinor{-s_z\tau_{\rm so} Q_1}{1}.
\end{equation}

\begin{figure}
	\includegraphics[width = 8.5 cm]{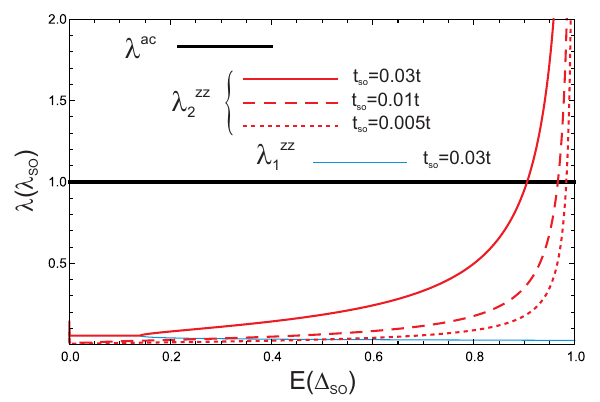}
	\caption{\label{Fig:EdgeStateWidth} (Color Online) Comparison of the topological edge state width of armchair ($\lambda^{\rm ac}$, black thick line) and zigzag ribbons ($\lambda^{\rm zz}_1$, blue thin line and $\lambda^{\rm zz}_2$, red lines) as a function of Fermi energy. The dominant $\lambda^{\rm zz}_2$ contribution in the zigzag case is also plotted for different values of $t_{\rm so}/t$ ($0.03$, $0.01$, $0.005$ for solid, dashed and dotted red lines, respectively).}
\end{figure}

It is interesting to note that one needs a superposition of two plane waves in order to satisfy the boundary condition [see Eq. (\ref{eq:ansatz_zigzag})], both for the zigzag and the armchair ribbon. In the zigzag case we need a superposition of two plane waves with different decay lengths in the transverse direction, see Eqs. (\ref{lambdaZZ}) and (\ref{eq:phivalues}). At low energies and for small values of the SO coupling, the contribution $\phi_2$ in Eqs.~ (\ref{eq:phivalues}) is clearly dominating. Furthermore, the SO coupling $\tau_{\rm so}$ only contributes a second order correction to $\phi_2$. Thus, the edge state width for zigzag ribbons is only weakly dependent on the exact value of the SO coupling. On the other hand, it is rather strongly dependent on the wave vector $q_x$ (through $Q_c$), and thus on energy. In Fig.~\ref{Fig:EdgeStateWidth} we have plotted the edge state widths $\lambda$ [given in Eqs. (\ref{lambdaAC}), (\ref{lambdaZZ}) and (\ref{eq:phivalues})] for armchair and zigzag orientations, normalised to $\lambda_{\rm so}$, as a function of the Fermi energy $E$ (normalised to $\Delta_{\rm so}$). The black thick line corresponds to $\lambda^{\rm ac}$, whereas the blue thin one corresponds to $\lambda^{\rm zz}_1$ and red ones to $\lambda^{\rm zz}_2$. Except for energy values very close to the SO gap, the width of the zigzag edge state is considerably smaller than the armchair one. This is precisely what we found in Fig.~\ref{Fig:chargedensity}. However, for energies close to $\Delta_{\rm so}$, the zigzag edge state width, $\lambda^{\rm zz}_2$ in particular, increases dramatically, as is also evident in Fig.~\ref{Fig:chargedensity}. For these energies, it also becomes dependent on the SO coupling strength, as shown in Fig. \ref{Fig:EdgeStateWidth} with the red solid, dashed and dotted lines. 

To understand the perfect transmission found in zigzag ribbons, first consider a ribbon without SO coupling. In that case, one can perform the low energy approximation around the two inequivalent $K_{1,2}$ points in the same fashion as with armchair terminated ones (see previous subsection). In the valley isotropic basis, the isospin is then simply given by $|\tau\rangle^l=(1,0)$ for a left moving particle and $|\tau\rangle^u=(0,1)$ for a right moving one at the lower and upper boundaries. Edge states along zigzag boundaries thus belong to opposite (orthogonal) valleys for opposite propagation directions. By turning on the SO coupling, the low energy dispersionless bands evolve into dispersive helical bands crossing at the $M$ point, as shown in Fig. \ref{Fig:system}(b), and their isospin (i.e., the valley polarization) is not well defined anymore. The latter bands are no longer separated by a large momentum, and one might therefore think that the potential step, even if it is smooth, could induce scattering between these bands. The key point now is that for a fixed spin, these two bands belong to opposite boundaries in the left region, so no large momentum can connect them. Scattering between these bands is only possible if the wavefunction is able to spread through the bulk of the ribbon. When this happens, to the right of the junction, the Fermi energy of the quasiparticles lies above the bulk gap where the high energy modes in this region have again a well defined valley polarization. Hence, all modes injected from a left moving helical edge state will belong to the, say, $K_1$ valley and not to the $K_2$ one. The backscattering channel, that is derived from the $K_2$ valley, will therefore remain inaccessible.

\begin{figure} 
\includegraphics[width=8cm]{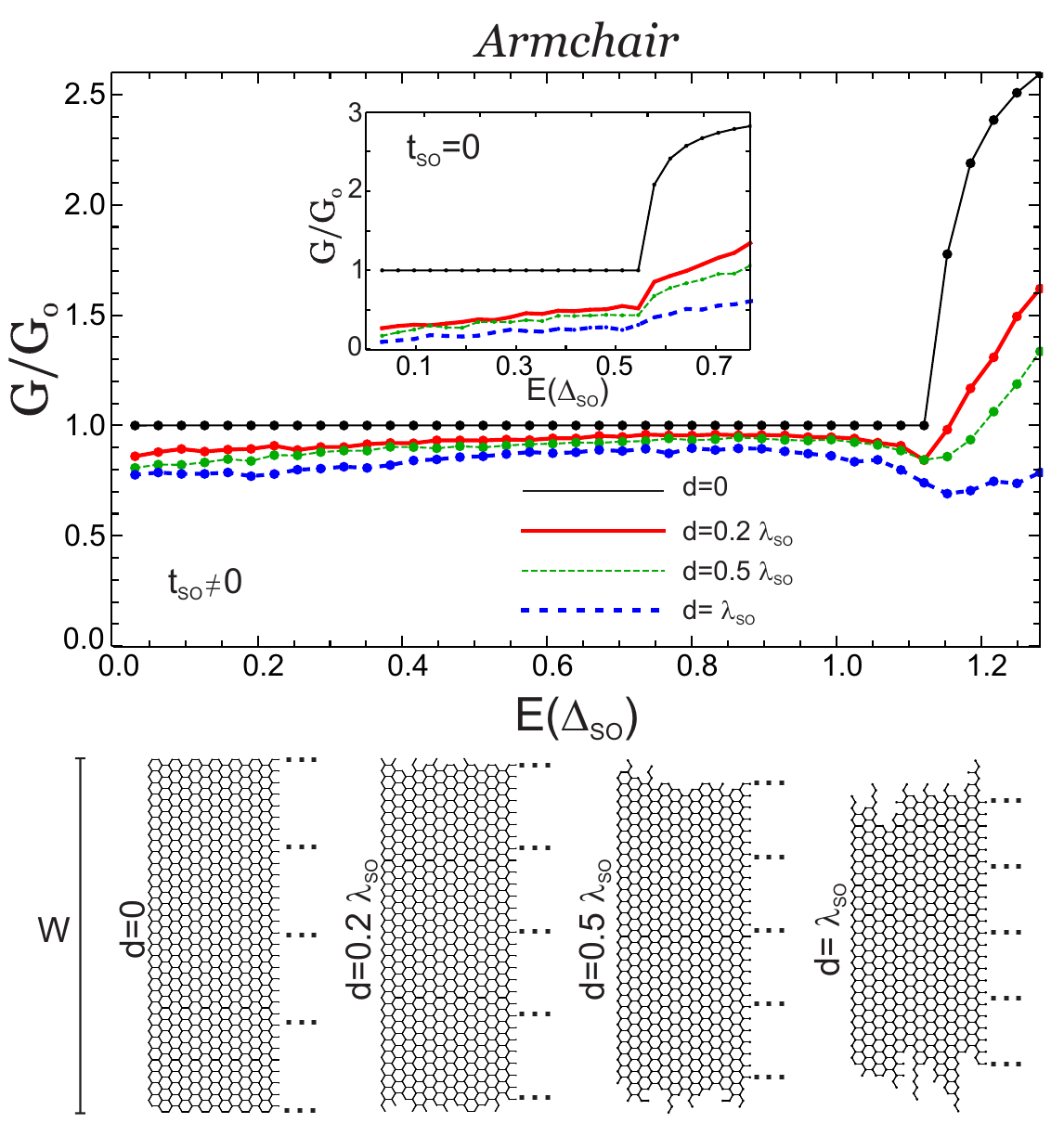}
\caption{\label{Fig:TRoughness-Armchair}(Color online) Upper panel: Conductance vs Fermi energy for a QSH insulator/metal junction in an armchair ribbon with edge roughness of different depths $d$.  The length of the disordered region is $L=10.4\lambda_{\rm so}$. Every data point is obtained by averaging over $256$ disorder configurations. Other parameters are as in Fig.~\ref{Fig:Tclean}. In the inset the same conductance is shown in the absence of spin orbit coupling. Lower panel: Ribbon geometry (finite section) for various roughness depths.
}
\end{figure}

\begin{figure} 
\includegraphics[width=8cm]{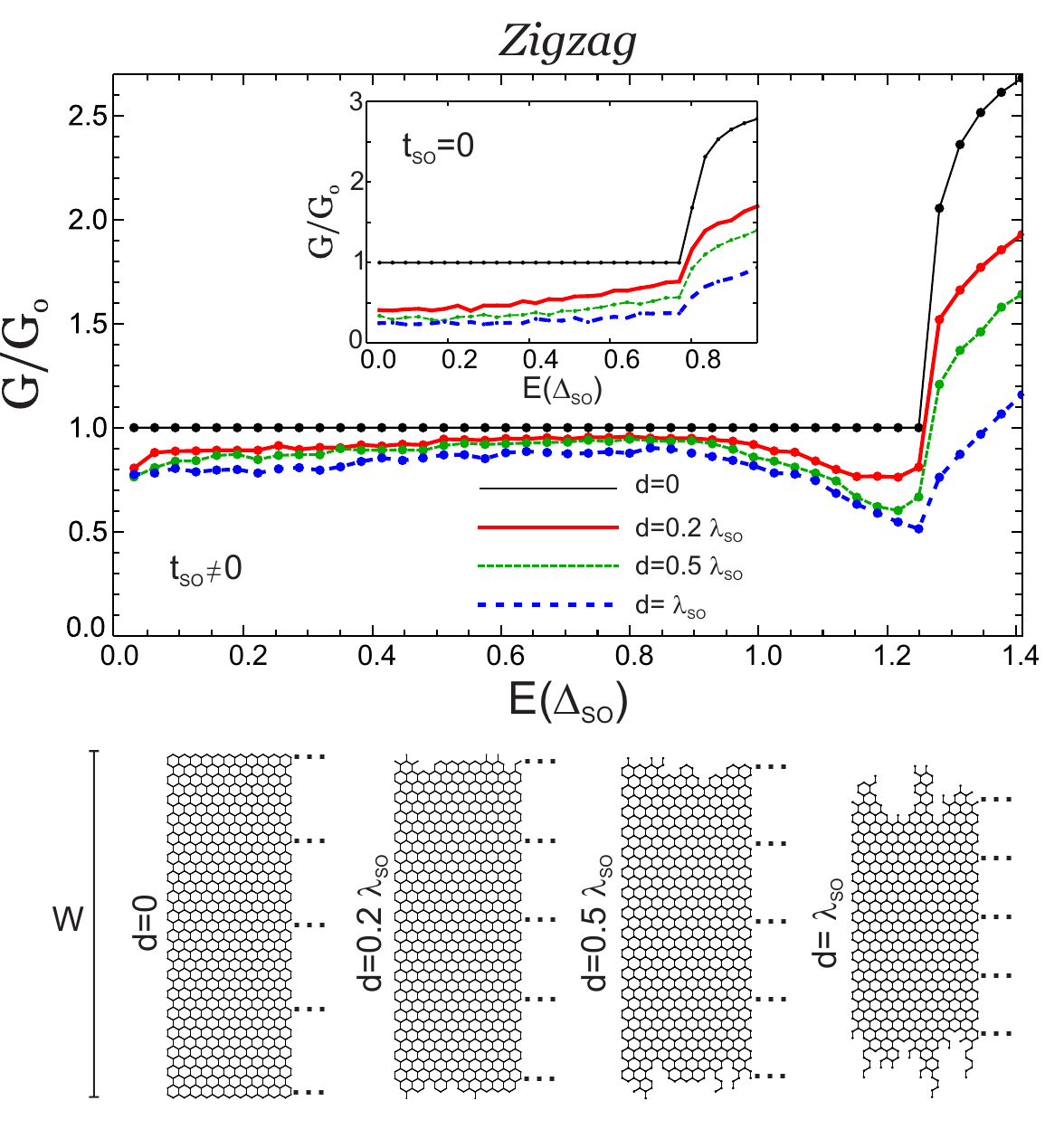}
\caption{\label{Fig:TRoughness-Zigzag}(Color online) The same as Fig. \ref{Fig:TRoughness-Armchair} but for a zigzag terminated ribbon.
}
\end{figure}

\section{Influence of edge roughness}\label{Sec:Disorder}
For a clean ribbon, perfect transmission was due to the orthogonality between equal-spin left- and right-moving edge channels, combined with valley-preserving scattering throughout all the ribbon. Disorder in general will break the orthogonality between topological edge channels and a decrease in the transmission through a QSH insulator/metal junction is thus to be expected. However, even if the disorder conserves the orthogonality of the topological channels, e.g., when disorder is only present in the right (metallic) region of the junction, it will in general induce inter-valley scattering which will connect counterpropagating edge channels through the bulk of the metallic region. Note that there are no topological edge states running along the junction, since this would require that both sides of the junction be bulk insulators with different values of the $Z_2$ topological invariant.

To analyze the influence of disorder in more detail we have considered rough edge boundaries \cite{Mucciolo2009}, which break the orthogonality between topological edge states. We model rough edges by randomly removing carbon atoms up to a depth $d$ (in units of $a$) from the ribbon's boundaries. In particular, we model a rough edge with a function of the form
\begin{equation}
	f(x) = y_b + \sum_{n=1}^{N/2} \frac{A_n d}{\sqrt{N}} \sin \left( \frac{2 \pi n}{N} x - \varphi_n \right).
\end{equation}
Here $y_b$ is $W-d/2$ for the upper boundary and $d/2$ for the lower one. The rough ribbon is obtained by removing all the carbon atoms in positions $(x,y)$ for which $y > f(x)$ ($y < f(x)$) at the upper (lower) boundary. The number of modes $N$ in the Fourier series is chosen to be of the order of the number of atoms along the length of the ribbon. The amplitudes $A_n$ and phases $\varphi_n$ are random numbers between $0$ and $1$ or $-\pi$ and $\pi$, respectively. Note that this function generates a boundary profile that varies strongly at the scale of the lattice spacing $a$ and therefore is a source of strong inter-valley scattering. Examples of ribbons with different roughness depths are sketched in the lower panels of Figs. \ref{Fig:TRoughness-Zigzag} and \ref{Fig:TRoughness-Armchair}, where sections of width $W=54a$ (as used in the calculations) are shown.

The conductance of a QSH insulator/metal junction in the presence of edge roughness is depicted in Figs.~\ref{Fig:TRoughness-Zigzag} and~\ref{Fig:TRoughness-Armchair} for armchair and zigzag ribbons, respectively. Results for intermediate ribbons are similar to one or the other (depending on the crystallographic orientation) and are not shown. All conductance points in these plots are obtained by averaging over $256$ disorder realizations. The non-perfect transmission of the topological edge states through the junction is manifested by a reduction of the first conductance plateau. This reduction is however much less pronounced than the conductance decrease in a disordered n-n (or p-p) junction in the absence of SO coupling (and thus without topological edge states in the left region), as shown in the insets of Figs.~ \ref{Fig:TRoughness-Armchair} and \ref{Fig:TRoughness-Zigzag}. 

\begin{figure} 
\includegraphics[width=7.5cm]{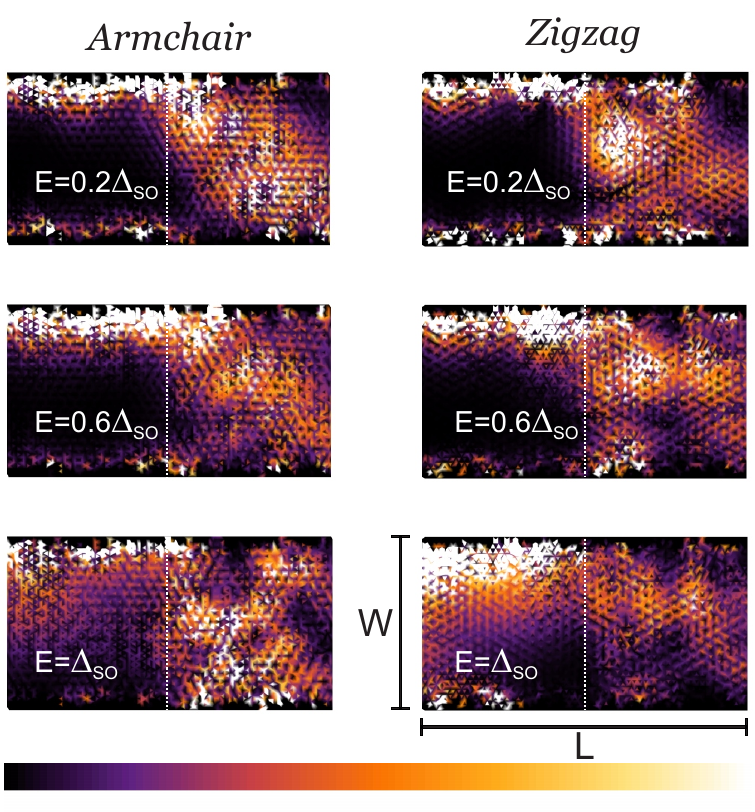}
\caption{\label{Fig:chargedensity-Roughness}(Color online) Numerical results for the charge density across the same potential step as in Fig. \ref{Fig:chargedensity}, but in the presence of edge disorder of depth $d=0.5\lambda_{\rm so}$. The topological edge channel from the upper left boundary is transmitted into bulk normal channels to the right but can now be backscattered towards the lower left boundary. Only one type of spin current is considered (a similar contribution with the opposite spin, not shown, is incoming from the lower boundary). 
}
\end{figure}

This increased protection against backscattering can be understood from plots of the charge density in the sample, as depicted in Fig.~\ref{Fig:chargedensity-Roughness}. In the presence of edge roughness, the spin-polarized edge state incoming from the upper boundary does not follow a straight line anymore (like in the flakes of Fig. \ref{Fig:chargedensity}) but its wavefuction accommodates to the roughened boundary. Therefore, unless $d$ is so large that upper and lower topological edge state wavefunctions can overlap, there are still no available intra-boundary backscattering channels. Only a single reflection channel is open, namely, to cross to the topological edge state at the lower boundary. This is radically different in a ribbon without intrinsic SO coupling, where multiple transmission and reflection channels are available throughout the bulk and disorder will have a bigger impact. The previous argument explains the small sensitivity of the conductance on roughness depth $d$ for energies below $\Delta_{\rm so}$ in Figs.~\ref{Fig:TRoughness-Armchair} and \ref{Fig:TRoughness-Zigzag}. However, when the edge state width $\lambda$ becomes comparable to the ribbon's effective width (given approximately by $W-2d$), direct inter-boundary backscattering can take place in the left region due to the overlap of counter-propagating (non-orthogonal) edge channels, significantly reducing the transmission probability. This happens for zigzag ribbons with energies close to the bandgap edge, where the edge state width diverges (see Fig.~\ref{Fig:EdgeStateWidth}), which manifests itself as a conductance dip in Fig.~\ref{Fig:TRoughness-Zigzag}. For armchair ribbons, this dip does not appear as the edge state width is independent of energy and much smaller than the ribbon width for the parameters we have chosen.

\begin{figure}
\includegraphics[width=8cm]{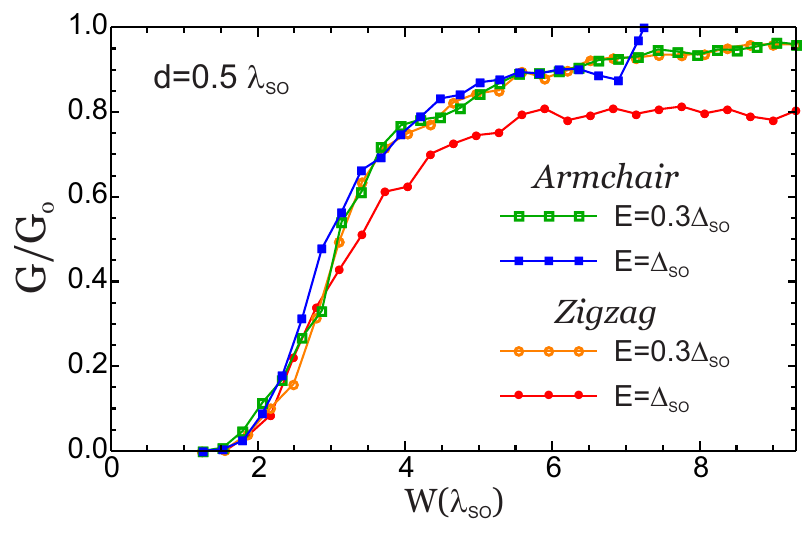}
\caption{\label{Fig:WyTRoughness} (Color online) Conductance for a QSH insulator/metal n-n junction as a function of the ribbon's width $W$. Edge disorder is present with roughness depth $d=0.5\lambda_{\rm so}$ and each conductance point is averaged over 180 configurations. Other parameters are as in Figs.~\ref{Fig:TRoughness-Armchair} and~\ref{Fig:TRoughness-Zigzag}. Zigzag and armchair ribbons with energies either close to the Dirac point or to the topological bandgap edge are compared.
}
\end{figure}

In Fig.~\ref{Fig:WyTRoughness} we plot the conductance through a QSH insulator/metal junction as a function of the ribbon's width $W$ in the presence of edge roughness  with a depth $d=0.5\lambda_{\rm so}$. The disordered region has a length $L= 10.4\lambda_{\rm so}$, and the potential step parameters are $U=1.6\Delta_{\rm so}$ and $\delta= 0.5\lambda_{\rm so}$. Zigzag and armchair ribbons are considered, and energies either close to the Dirac point or close to the SO induced gap are compared. For all parameters, we clearly see that the conductance increases with $W$, so that transmission approaches one in the thermodynamic limit. This can be understood by realizing that, as the ribbon width is increased, the number of available forward scattering channels grows while the available backscattering channels remains one. Note that for the potential step parameters selected, all curves behave similarly except for the one of the zigzag terminated ribbon for energies close to the gap. The slower conductance increase in that case is a result of the large overlap of the opposite boundary topological edge states, as discussed previously. Although the junction used for the calculations in Fig.~\ref{Fig:WyTRoughness} is a n-n junction, we have found similar results for p-p and n-p (p-n) junctions and conclude that conductance recovery with increasing ribbon's width is a general trend.

\section{Conclusions}\label{Sec:Conclusions}
In this work we investigate electron transport through QSH insulator/metal junctions in graphene ribbons. These kind of studies are of special relevance for on-going and future experiments and devices implying topological edge states, since transport through such systems will be ultimately limited by the contact resistance. Here, QSH insulator/metal junctions are created in graphene ribbons by means of an electrostatic potential step. We have found that for clean ballistic ribbons (with arbitrary crystallographic orientation), the topological edge states are transmitted perfectly through such a junction, independent of the ribbon's width, the parameters of the potential step and the value of the intrinsic SO coupling, as long as inter-valley scattering is not present. This has been explained by proving the isospin orthogonality between forward and backward edge channels analytically, a mechanism that is analog to the Klein tunneling. We have also found that the edge state width only depends on the SO coupling strength for armchair ribbons but is much smaller, and in addition strongly energy dependent, for zigzag ribbons. In the presence of edge roughness, the orthogonality, and thereby the perfect transmission, is destroyed and the upper (lower) edge channel can be reflected to the lower (upper) one through the metallic region where the particle's wave function spans the whole ribbon's width. However, even in the presence of strong disorder, reflection from a QSH insulator/metal junction is small as compared to a similar ribbon in the absence of intrinsic SO coupling. This protection can be understood as a SO proximity effect~\cite{Prada2010b} since to the left of the junction the edge states are still topologically protected (as long as time-reversal symmetry breaking disorder is absent). Finally, we found that perfect transmission is gradually restored in the presence of disorder as the ribbon's width increases, since the number of available forward-scattering channels grows. In the present work, the QSH insulator/metal junction was realized by means of a constant SO coupling and an electrostatic potential step. An analogous junction could be achieved without modulating the doping but in the presence of a SO potential step if the intrinsic SO coupling could be externally tuned \cite{Pereg-Barnea2010} across the ribbon. We also analyzed this scenario numerically and found similar results for metallic ribbons.
Finally, we would like to mention that the perfect transmission we find for graphene QSH insulator systems in contact with a metallic phase might be rather specific of graphene, since it is ultimately based on the particular symmetry properties of its wave function. For this reason, we conjecture that our conclusions for the clean case will not apply to other QSH systems like for instance HgTe/CdTe quantum wells.

\section*{Acknowledgments}
E.P. acknowledges the support of the CSIC JAE-Doc program and the Spanish Ministry of Science and Innovation through Grant No. FIS2009-08744.

\end{document}